\documentclass[5p]{elsarticle}

\usepackage{xparse}
\usepackage{amsmath}
\usepackage{amsfonts}
\usepackage{amssymb}
\usepackage{graphicx}
\usepackage{epstopdf}
\usepackage{url,ifthen}
\usepackage{caption}
\usepackage{subcaption}
\usepackage{nopageno}
\usepackage[dvipsnames,usenames]{color}
\usepackage[colorlinks, linkcolor=Blue, filecolor =Red,citecolor=RoyalPurple]{hyperref}

\newtheorem{comment}{Comment}
\def\R{\mathbb{R}}

\def\eqdef{\ensuremath{:=}}

\def\qbarAgg{\ensuremath{{\bar{q}_\mathrm{agg}}}}

\def\hatqbarAgg{\ensuremath{{\hat{\bar{q}}_\mathrm{agg}}}}
\def\dotqbarAgg{\ensuremath{{\dot{\bar{q}}_\mathrm{agg}}}}

\newlength{\noteWidth}
\long\def\notes#1{\ifinner
	{\footnotesize #1}
	\else
	\marginpar{\parbox[t]{\noteWidth}{\raggedright\footnotesize #1}}
	\fi\typeout{#1}}
\def\pb#1{\notes{pb: {\color{red}{#1}}      }}

\ExplSyntaxOn
\cs_new:Nn \pb_prop_gset_bykeys:Nn
{
	\prop_clear:N \l__pb_temp_prop
	\keys_set:nn { pb/propbykey } { #2 }
	\prop_gset_eq:NN #1 \l__pb_temp_prop
}
\cs_generate_variant:Nn \pb_prop_gset_bykeys:Nn { c }

\keys_define:nn { pb/propbykey }
{
	unknown .code:n = \prop_put:Nxn \l__pb_temp_prop { \l_keys_key_tl } { #1 }
}

\cs_generate_variant:Nn \prop_put:Nnn { Nx }

\NewDocumentCommand{\setupcollaborator}{mm}
{
	\prop_new:c { g_collaborator_#1_prop }
	\pb_prop_gset_bykeys:cn { g_collaborator_#1_prop } { #2 }
}

\NewDocumentCommand{\selectcollaborator}{m}
{
	\prop_map_inline:cn { g_collaborator_#1_prop }
	{
		\tl_set:cn { ##1 } { ##2 }
	}
}
\ExplSyntaxOff

\setupcollaborator{PBmac}
{
	NAME=Prabir Mac , laptop or mac mini,
	DiCEbibPATH=/Users/pbarooah/Dropbox/DiCElab_bib,
	ACbibPATH=/Users/pbarooah/Dropbox/austin_bib,
	NSRbibPATH=/Users/pbarooah/Dropbox/NSRbib,
    JBbibPATH =/Users/pbarooah/Dropbox/bibs,
}
\setupcollaborator{GZlinux}
{
	NAME=GZ , on his lab computer,
	DiCEbibPATH=../../../DiCElab_bib,
	ACbibPATH=../../../austin_bib,
	JBbibPATH = ../../../bibs,
}

\setupcollaborator{NRwindows}
{
	NAME=NR on his windows computer,
	DiCEbibPATH=../../../../../DiCElab_bib,
	ACbibPATH=../../austin_bib,
}
\setupcollaborator{NRlinux}
{
	NAME=NR on his linux computer,
	DiCEbibPATH=../../../../../DiCElab_bib,
	NSRbibPATH=../../../../../NSRbib,
}

\selectcollaborator{GZlinux}

\begin{document}
	\begin{frontmatter}
	\title{Aggregation and data driven identification of building thermal dynamic model and unmeasured disturbance}
	
	\author[mymainaddress]{Zhong Guo} 
	\author[mymainaddress]{Austin R. Coffman\corref{cor1}}
	\ead{bubbaroney@ufl.edu}	
	\author[mysecondaryaddress]{Jeffrey Munk}
	\author[mysecondaryaddress]{Piljae Im}
	\author[mysecondaryaddress]{Teja Kuruganti}
	\author[mymainaddress]{Prabir Barooah}
		
	\cortext[cor1]{Corresponding author}

	\address[mymainaddress]{Department of Mechanical and Aerospace Engineering, University of Florida, Gainesville, FL 32611, USA}
	\address[mysecondaryaddress]{Oak Ridge National Laboratory, Oak Ridge, TN 37830, USA }
	\begin{abstract}
          An aggregate model is a single-zone equivalent of a multi-zone building, and is useful for many purposes, including model based control of large heating, ventilation and air conditioning (HVAC) equipment. This paper deals with the problem of simultaneously identifying an aggregate thermal dynamic model and unknown disturbances from input-output data. The unknown disturbance is a key challenge since it is not measurable but non-negligible. We first present a principled method to aggregate a multi-zone building model into a single zone model, and show the aggregation is not as trivial as it has been assumed in the prior art. We then provide a method to identify the parameters of the model and the unknown disturbance for this aggregate (single-zone) model. Finally, we test our proposed identification algorithm to data collected from a multi-zone building testbed in Oak Ridge National Laboratory. A key insight provided by the aggregation method allows us to recognize under what conditions the estimation of the disturbance signal will be necessarily poor and uncertain, even in the case of a specially designed test in which the disturbances affecting each zone are known (as the case of our experimental testbed). This insight is used to provide a heuristic that can be used to assess when the identification results are likely to have high or low accuracy.
	\end{abstract}
\end{frontmatter}
\section{Introduction}
A dynamic model of a building zone temperature is useful in several applications involving heating, ventilation, and air conditioning (HVAC) systems, such as model-based control for improving indoor climate and reducing energy use~\cite{afram2014theory,Ma:2012}, limiting peak demand~\cite{Morris:94,Sun:2013}, or providing ancillary services to the power grid~\cite{haomidbarmey:2012,Lin:2015}. To be used in a control algorithm, especially real-time control, the model should be of low order. A common way to achieve this is estimating the model from input-output measurements~\cite{Ma:2012,haomidbarmey:2012,Lin:2015,WANG2019109405,Fux:2014,Bacher:2011,Wang:2006,Andersen:2000,Kim_Braun:2016,cofbar:2018,ZengSimultaneousHPB:2018}.

There are many different model structures for modeling the thermal dynamics of a building zone. The Resistance-Capacitance (RC) network is a commonly used model structure; use of such models goes back a long way~\cite{PahwaThesis:1983}. There  is a rich literature on identification of the parameters of a RC network model for single zone commercial buildings~\cite{Fux:2014,Bacher:2011,Wang:2006,Andersen:2000}. For a single zone model, the output is the zone temperature, while the inputs are the heating (or cooling) rate injected by the HVAC system, ambient air temperature, solar irradiance and the internal heat load. All inputs can be easily quantified except the internal heat load. The internal heat load is a sum of all sources of heat inside the building, such as occupants' metabolism, and lights and appliances used by occupants. The internal heat load is an exogenous disturbance that is typically unknown and not measurable. Since this disturbance can be comparable in magnitude to the cooling provided by the HVAC system,  it poses a challenge in model identification from measured input-output data~\cite{Kim_Braun:2016}. Work on identifying the model in spite of the presence of this unknown disturbance is quite recent~\cite{Kim_Braun:2016,cofbar:2018,ZengSimultaneousHPB:2018}. Earlier works have mostly ignored this disturbance; see~\cite{cofbar:2018} for a discussion of earlier approaches.

More challenges arise when we extend modeling and/or identification approaches from single-zone to multi-zone buildings~\cite{ZengIdentificationCPHS:2018}. The multi-zone model is much more complicated since there are more inputs, outputs, parameters, and states. In particular, despite the internal heat load for a single zone is already hard to identify, there can be as many internal heat loads (unknown disturbances) as there are zones. The identification of a multi-zone model is thus more challenging than that of a single zone model. Many works on model based control for multi-zone buildings instead use a ``single-zone equivalent" model of the building, in which the average building temperature (averaged over the zones) and the sum of zone-level inputs are used~\cite{ThibaultPean:2018,Ma:2014,patel2018economic}. Such a model is useful in determining building-level control commands for large equipment such as a chiller or an air handling unit that serves a multi-zone building.

In this paper we address the problem of identification of such an aggregate model. A principled approach to aggregating a multi-zone building into a single zone equivalent is proposed first. We then provide a method to identify thermal parameters and the unknown disturbance for this aggregate (single-zone) model. Lastly, we test our proposed identification algorithm to data generated from simulation and data collected from a multi-zone building testbed at the Oak Ridge National Laboratory in which disturbances are known. Evaluation from both datasets shows that the proposed method performs quite well. The insight on the difference between the average internal heat loads and the aggregate heat load is useful in evaluating the estimation results.

\subsection{Contribution over prior art}

The concept of an aggregate model is not new in building modeling and control literature. Many Model Predictive Control (MPC) formulations use an aggregate model to determine building-level control commands~\cite{patel2018economic,energies:MPC}. There is in fact a plethora of works that reduces a model of a multi-zone building to that of a single ``aggregate'' zone. A common practice is averaging/summing the inputs and outputs over all zones to form a single set of input and output signals. These signals are then assumed to be related by an arbitrary RC network model, which is then set up as a system identification problem to determine the parameters of the chosen RC network structure~\cite{Fux:2014,Bacher:2011,Wang:2006}. The difference between our work and those methods just described, is that we \emph{first} construct our aggregate model through analysis of the multi-zone model. Other works, e.g., \cite{Deng:2014_Automatica,patel2018economic}, aggregate a multi-zone model into a single zone model using knowledge of each zone's thermal parameters and inter-zone interactions, which are usually unknown, limiting their applicability.

Our aggregate model is derived from the RC network models of individual zones and constructed definitions of aggregate input and output signals. The aggregate input and output signals are solely functions of the same signals of each zone and the derived aggregate model has time invariant parameters. Therefore the aggregate model lends itself well to system identification approaches.

There are four contributions of our work. The first contribution is a principled method of aggregating a high-dimensional multi-zone model into a low-order single zone equivalent model which we called an ``\emph{aggregate model}". This clears the ambiguity in the definition of inputs and outputs of the aggregate model. An outcome of this method is an insight into the appropriate definition of the internal heat load for the aggregate model, which we call the \emph{aggregate internal heat load}. This aggregate internal heat load turns out \emph{not} to be the average of the internal heat loads of individual zones. Rather, it is a non-trivial function of many signals.  Our second contribution is a novel method to estimate both model parameters and the aggregate internal heat load from measurable input-output data. This identification method is inspired by the Kalman-filter based method in~\cite{cofbar:2018}. The difference is that the method proposed here allows for the incorporation of prior knowledge in the form of constraints (such as non-negativity of the internal heat load) while that in~\cite{cofbar:2018} does not. The third contribution is a heuristic based on the insights obtained during aggregation that allows one to predict - based purely on the data used for identification - when the estimated aggregate internal heat load is likely to be accurate (or not). The fourth contribution is assessment of the proposed method with data collected from a special test building in which the internal heat loads are carefully controlled and thus made measurable. This building is a unique facility in which it is possible to measure the occupant induced heat loads: space heaters in each zone are controlled to mimic occupant behavior. To the best of our knowledge, no other works have reported results from such experiments, including those who have evaluated methods on building data. The lack of ground truth makes such evaluations challenging to assess. In contrast, by performing special experiments in this building so that the ground truth is known, the results can be assessed far more clearly.

A preliminary version of this paper is presented in~\cite{GuoIdentification:CDC:19}. While \cite{GuoIdentification:CDC:19} only presented simulation evaluation, this paper  presents evaluation of the proposed method with data from a real building. The second contribution over~\cite{GuoIdentification:CDC:19} is the heuristic to predict when the identification results are likely to be accurate.

The rest of the paper is organized as follows: Section~\ref{sec:Two} describes the process of modeling the thermal dynamics of multi-zone buildings. Section~\ref{sec:Three} formulates the identification problem and proposes our identification method along with a heuristic to predict the accuracy of estimated disturbance. Section~\ref{sec:Four} shows the identification results for: (i) simulation data and (ii) data collected from a real building. We conclude in Section~\ref{sec:conclusion}.

\section{Building Thermal Modeling Structure}\label{sec:Two}
\subsection{Single-zone Model}
\begin{figure}
		\centering
		\includegraphics[width=0.7\columnwidth]{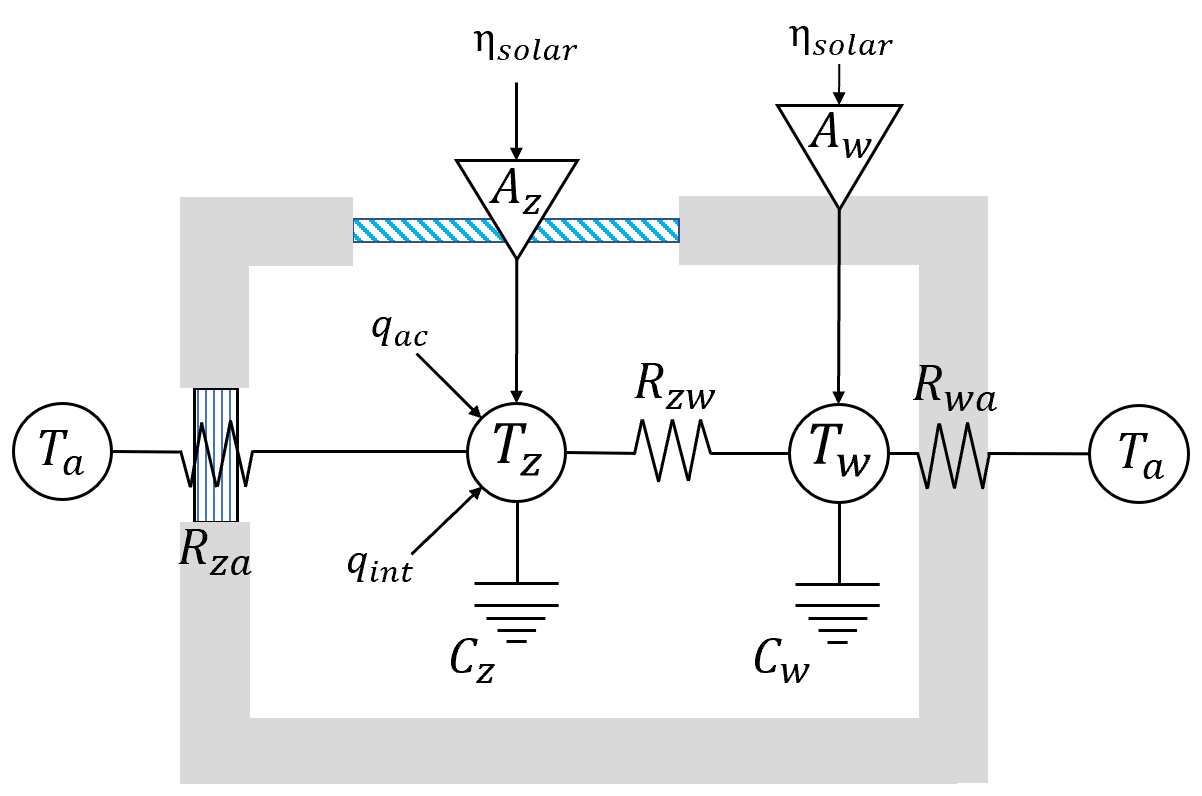}
		\caption{2R2C network on top of a single zone building.}
		\label{fig:Ind_RC_structure}
	\end{figure}

We start with a single zone and its model. A floor plan for a single zone is presented in Figure~\ref{fig:Ind_RC_structure}. A Resistance-Capacitance (RC) network model, particularly a 2R2C model, is overlaid on the floor plan. The 2R2C model refers to the following coupled differential equations that describe the evolution of temperature of the zone, $T_z$, in response to various inputs:
\begin{align}
\dot{T}_z(t) &= \frac{T_a(t) - T_z(t)}{R_{za}C_z} + \frac{1}{C_z}(q_{int}(t) - q_{ac}(t))
\nonumber\\&+ \frac{T_w(t) - T_z(t)}{R_{zw}C_z} + \frac{A_{z}}{C_z} \eta_{solar}(t),\label{eq:Ind_model_Tz}
\\\dot{T}_w(t) &= \frac{T_z(t) - T_w(t)}{R_{zw}C_w} + \frac{T_a(t) - T_w(t)}{R_{wa}C_w} + \frac{ A_{w}}{C_w} \eta_{solar}(t),\label{eq:Ind_model_Tw}
\end{align}
where $T_w$ is a fictitious state (temperature of the wall, if you must), $q_{ac}$ is the cooling load due to the HVAC system, and $\eta_{solar}$, $T_{a},$ and $q_{int}$ are non-controllable inputs, which are the solar irradiance, ambient (outside) temperature, and internal heat load, respectively. The parameters in this model are $\{R_{za},C_z,R_{zw},C_w,A_z,A_w\}$.

	\subsection{Aggregate Multi-zone Model}\label{sec:aggDeriv} 
\begin{figure}
	\centering
	\includegraphics[width=0.65\columnwidth]{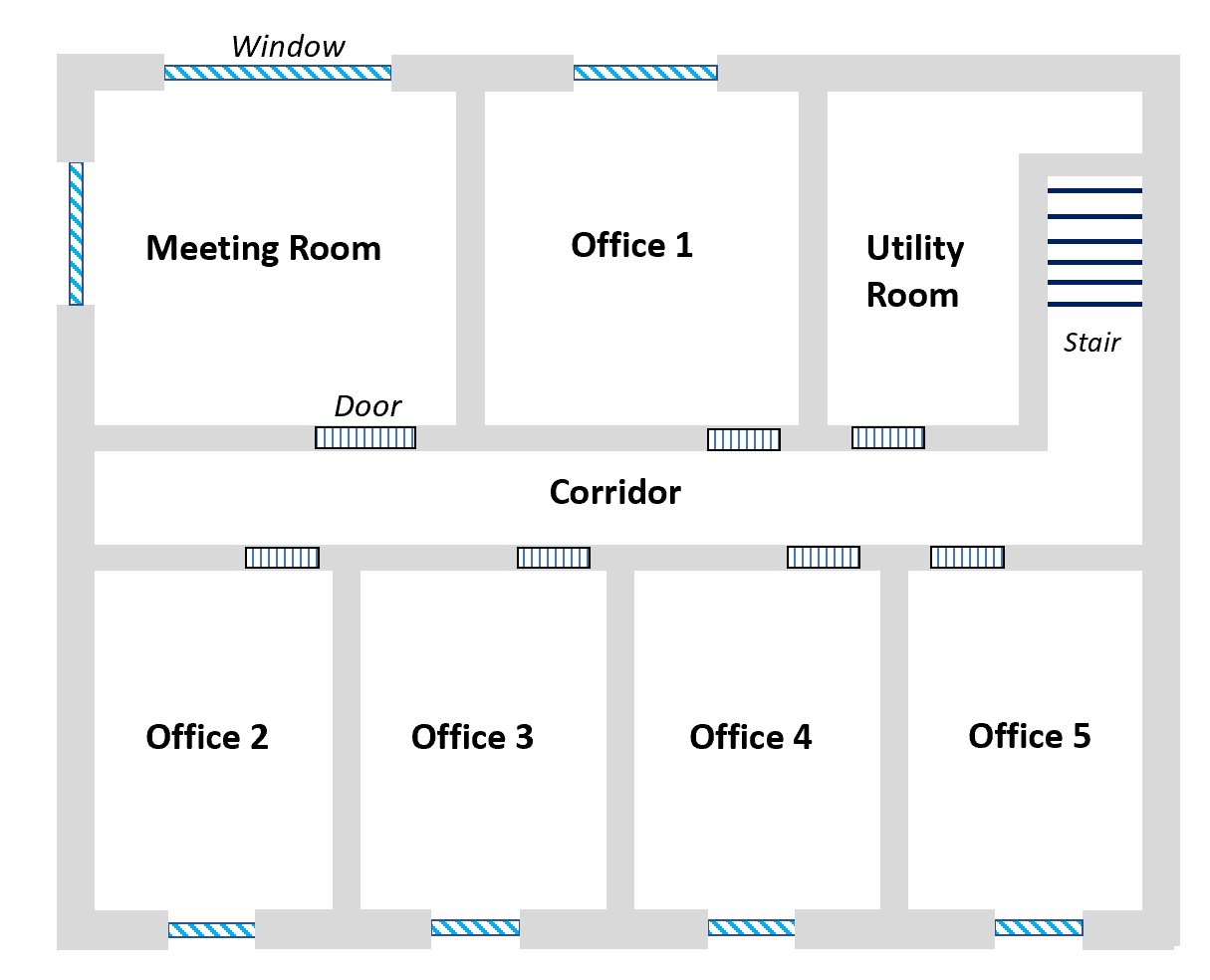}
	\caption{Example floor plan for a multi-zone commercial building.}
	\label{fig:floor_plan}
	\end{figure}

	Here we describe how a model of a multi-zone building can be aggregated into a single-zone model of the type described in the previous section. An example of multi-zone building is shown is Figure~\ref{fig:floor_plan}. In the sequel, we call the resulting single-zone model as the \emph{aggregate model} for the multi-zone building. 

	To start the derivation, let $N_z$ be the number of zones in the building, and each zone is modeled by a 2R2C network model using~\eqref{eq:Ind_model_Tz}-\eqref{eq:Ind_model_Tw}. \emph{We now define the average input/output quantities as follows:}
	\begin{align}
	&\bar{T}_{p}(t) \triangleq \frac{1}{N_{z}}\sum_{j=1}^{N_{z}} T_{p}^j(t), \quad \bar{q}_{ac}(t) \triangleq \frac{1}{N_z}\sum_{j=1}^{N_{z}} q_{ac}^j(t), \label{eq:barVariables1}\\ 
	&\bar{q}_{int}(t) \triangleq \frac{1}{N_z}\sum_{j=1}^{N_{z}} q_{int}^j(t),\  \bar{\eta}_{solar}(t) \triangleq \frac{1}{N_z}\sum_{j=1}^{N_{z}} \eta_{solar}^j(t),  \label{eq:barVariables2}
	\end{align}
	where $T_p^j$ is the temperature of $p^{th}$ location for the $j^{th}$ zone. For example, $T_z^2$ represents the zone temperature of the 2nd zone. In the sequel, \emph{the bar is used to distinguish an average quantity}.

	By summing the $N_z$ individual thermal models~\eqref{eq:Ind_model_Tz}-\eqref{eq:Ind_model_Tw} over the $j$ index and dividing by the number of zones, we have:
	\begin{align} 
	\frac{1}{N_z}\sum_{j=1}^{N_{z}}\dot{T}_z^j(t) & =\nonumber \frac{1}{N_z}\sum_{j=1}^{N_{z}}\left(\frac{{T}_a^j(t) - {T}_z^j(t)}{R_{za}^jC_z^j} + \frac{T_w^j(t) - T_z^j(t)}{R_{zw}^jC_z^j}\right)
	\\ &+ \frac{1}{N_z}\sum_{j=1}^{N_{z}} \left(\frac{q_{int}^j(t)-q_{ac}^j(t)}{C_z^j} + \frac{A_{z}^j}{C_{z}^j} \eta_{solar}^j(t)  \right) \label{eq:average-Tz-multizone},
	\\
	\frac{1}{N_z}\sum_{j=1}^{N_{z}}\dot{T}_w^j(t) &=\nonumber \frac{1}{N_z}\sum_{j=1}^{N_{z}}\left(\frac{T_z^j(t) - T_w^j(t)}{R_{zw}^jC_w^j} +\frac{A_{w}^j}{C_w^j} \eta_{solar}^j(t) \right) \\ 
	&+\frac{1}{N_z} \sum_{j=1}^{N_{z}}\frac{T_a^j(t) - T_w^j(t)}{R_{wa}^jC_w^j}.\label{eq:average-Tw-multizone}
	\end{align} 

	Notice that above we have assumed that the zones have no thermal interactions with each other. We have done so only in the interest of saving space; later we comment on the changes if thermal interactions among zones exist. 
	
	Now we can construct the aggregate model by substituting definitions~\eqref{eq:barVariables1}-\eqref{eq:barVariables2} into equations~\eqref{eq:average-Tz-multizone}-\eqref{eq:average-Tw-multizone}. Two approaches are possible, leading to a time varying model or a time invariant model, respectively.
	\subsubsection{Time Varying Model}
	
	If we define a model from~\eqref{eq:average-Tz-multizone}-\eqref{eq:average-Tw-multizone} whose structure is exactly the same as that of the single-zone model~\eqref{eq:Ind_model_Tz}-\eqref{eq:Ind_model_Tw} in terms of the average signals defined in~\eqref{eq:barVariables1}-\eqref{eq:barVariables2}, the thermal model becomes:
	\begin{align}
	\dot{\bar{T}}_z(t) &= \frac{\bar{T}_a(t) - \bar{T}_z(t)}{\bar{\tau}_{za}(t)} - \frac{\bar{q}_{ac}(t)}{\bar{C}_z^{ac}(t)} + \frac{\bar{q}_{int}(t)}{\bar{C}_z^{int}(t)} \nonumber \\
	&+ \frac{\bar{T}_w(t) - \bar{T}_z(t)}{\bar{\tau}_{zw}(t)} + \bar{A}_{z}(t)\bar{\eta}_{solar}(t),\label{eq:time_varying_T_z}\\
	\dot{\bar{T}}_w(t) &= \frac{\bar{T}_a(t) - \bar{T}_w(t)}{\bar{\tau}_{wa}(t)} + \frac{{T}_z(t) - \bar{T}_w(t)}{\bar{\tau}_{wz}(t)} + \bar{A}_{w}(t)\bar{\eta}_{solar}(t),\label{eq:time_varying_T_w}
	\end{align}
	The expression for the thermal parameters are:
	\begin{align} 
	& \bar{\tau}_{pm}(t) \triangleq \frac{N_z(\bar{T}_m(t) - \bar{T}_p(t))}{\sum_{j=1}^{N_{z}}\frac{{T}_m^j(t) - {T}_p^j(t)}{R_{pm}^jC_p^j}} \label{eq:time_invar_para1},\\
	& \bar{C}_p^{m}(t) \triangleq \frac{N_z\bar{q}_{m}(t)}{\sum_{j=1}^{N_{z}}\frac{q_{m}^j(t)}{C_p^j}}\label{eq:time_invar_para2},\\
	&\bar{A}_{p}(t) \triangleq \frac{\sum_{j=1}^{N_{z}}\frac{A_p^j}{C_p^j}\eta_{solar}^j(t)} {N_z\bar{\eta}_{solar}(t)}\label{eq:time_invar_para3},
	\end{align}
	which are time varying.
	To save space, the sub/superscripts on the LHS values denote the relevant location (like the $x$ subscript in~\eqref{eq:barVariables1}).
	The right hand side of these equations become time-invariant only if either thermal parameters or input variables are completely homogeneous over the zone index $j$. Otherwise, these defined ``thermal parameters'' will be time-varying depending on each zone's individual inputs and states, which is not an appealing situation.
	
	\subsubsection{Time Invariant Aggregate Model}
	Since the thermal parameters and inputs for individual zones are unlikely to be homogeneous over each zone, we modify our interpretation of the average signals~\eqref{eq:barVariables1}-\eqref{eq:barVariables2} so that the aggregate model becomes a RC network model with time invariant thermal parameters. Doing so requires defining the following deviation variables:
	\begin{align} \label{eq:define_tilde}
	\tilde{T}_{p}^j &\triangleq T_{p}^j - \bar{T}_{p}, \quad \tilde{q}_{ac}^j \triangleq q_{ac}^j - \bar{q}_{ac},  \nonumber\\
	\tilde{q}_{int}^j &\triangleq q_{int}^j - \bar{q}_{int}, \quad \tilde{\eta}_{solar}^j \triangleq \eta_{solar}^j - \bar{\eta}_{solar},
	\end{align}
	where $p$ is the same as \eqref{eq:barVariables1}. The interpretation of \eqref{eq:define_tilde} is that all individual inputs and states can be represented as the average quantity (zonal average) plus some deviation. With some tedious algebra, it can be shown that with the help of these definitions, the ODEs~\eqref{eq:time_varying_T_z}-\eqref{eq:time_varying_T_w} can be transformed to:
	\begin{align}
	\dot{\bar{T}}_z(t)  = &\frac{\bar{T}_a(t)-\bar{T}_z(t)}{\bar{\tau}_{za}}  + \frac{\bar{T}_w(t)-\bar{T}_z(t)}{\bar{\tau}_{zw}}\\
	&+\frac{1}{\bar{C}_z}\left(\bar{q}_{int}(t) - \bar{q}_{ac}(t)\right) + {\bar{A}_{z}} \bar{\eta}_{solar}(t) +\tilde{w}_{z}(t), \nonumber\\
	\dot{\bar{T}}_w(t)  = &\frac{\bar{T}_a(t)-\bar{T}_w(t)}{\bar{\tau}_{wa}} + \frac{\bar{T}_z(t)-\bar{T}_w(t)}{\bar{\tau}_{wz}} \nonumber\\
	&+ \bar{A}_{w} \bar{\eta}_{solar}(t) +\tilde{w}_{w}(t),
	\end{align}
	where
	\begin{align}  \label{eq:aggParam}
	\bar{\tau}_{pm} \triangleq \frac{N_z}{\sum_{j=1}^{Nz} \frac{1}{R_{pm}^jC_p^j}},
	&\quad \bar{C}_p \triangleq \frac{N_z}{\sum_{j=1}^{Nz} \frac{1}{C_p^j}},
	&\bar{A}_{p} \triangleq \frac{\sum_{j=1}^{Nz} \frac{A_p^j}{C_p^j}}{N_z},	
	\end{align}
	\begin{align} \label{eq:wTildeZ}
	\tilde{w}_{z}(t) \triangleq& \frac{1}{N_z}\sum_{j=1}^{N_{z}}\frac{\tilde{T}^j_{a}(t)-\tilde{T}^j_{z}(t)}{R_{za}^{j}C_z^{j}} + \frac{1}{N_z}\sum_{j=1}^{N_{z}}\frac{\tilde{T}^j_{w}(t)}{R_{zw}^{j}C_z^{j}}  \nonumber\\ \nonumber
	& +\frac{1}{N_z}\sum_{j=1}^{N_{z}}\frac{\tilde{q}^j_{int}(t) - \tilde{q}^j_{ac}(t) + A_{z}^j\tilde{\eta}_{solar}^j(t) }{C_z^{j}}  \nonumber\\
	\end{align}
	\begin{align}\label{eq:wTildeW}
	\tilde{w}_{w}(t) \triangleq &\frac{1}{N_z}\sum_{j=1}^{N_{z}}\frac{\tilde{T}^j_{a}(t)-\tilde{T}^j_{w}(t)}{R_{wa}^{j}C_w^{j}} +\frac{1}{N_z}\sum_{j=1}^{N_{z}}\frac{\tilde{T}^j_{z}(t)}{R_{zw}^{j}C_w^{j}}  \nonumber\\ 
	&+ \frac{1}{N_z}\sum_{j=1}^{N_{z}} \frac{A_{w}^j}{C_w^j} \tilde{\eta}_{solar}^j(t).
	\end{align}
	The thermal parameters of the aggregate model are $\bar{\tau}_{xy}$, $\bar{C}_x$, and $\bar{A}_{x}$ and are time invariant.	The signals $\tilde{w}_z,\tilde{w}_w$ are additive time-varying terms that represent model mismatch due to the asynchronous inputs and states for each zone and wall, respectively. We term these additive terms as \emph{``aggregation errors''} and they will be zero only if all deviation variables are synchronous. \emph{We define the aggregate internal heat load as:}  
	\begin{align}\label{eq:def-q-agg}
	\qbarAgg(t) \triangleq \bar{q}_{int}(t) + \tilde{w}_z(t)\bar{C}_z,
	\end{align} 
	which leads to the final construction of the aggregate model:
	\begin{align}
	\dot{\bar{T}}_z(t)  = &\frac{\bar{T}_a(t)-\bar{T}_z(t)}{\bar{\tau}_{za}} +\frac{\bar{T}_w(t)-\bar{T}_z(t)}{\bar{\tau}_{zw}}+\bar{A}_{z} \bar{\eta}_{solar}(t), \nonumber\\
	& + \frac{1}{\bar{C}_z}(\qbarAgg(t) - \bar{q}_{ac}(t))\label{eq:timeInvAggModel_Tz}\\
	\dot{\bar{T}}_w(t)  = &\frac{\bar{T}_a(t)-\bar{T}_w(t)}{\bar{\tau}_{wa}}  + \frac{\bar{T}_z(t)-\bar{T}_w(t)}{\bar{\tau}_{wz}} \nonumber\\
	&+\bar{A}_{w} \bar{\eta}_{solar}(t)+ \tilde{w}_{w}(t).\label{eq:timeInvAggModel_Tw}
	\end{align}
	\begin{comment} \label{com:commentOne}
	The derived model~\eqref{eq:timeInvAggModel_Tz}-\eqref{eq:timeInvAggModel_Tw} represents the time evolution for ``average'' quantities of building zones. That is, if data is collected from individual zones in a building and is aggregated according to~\eqref{eq:barVariables1}-\eqref{eq:barVariables2}, then equation~\eqref{eq:timeInvAggModel_Tz}-\eqref{eq:timeInvAggModel_Tw} describes how these states should change over time. Furthermore, if this average data is utilized in system identification of an equivalent single-zone RC-network model, then~\eqref{eq:aggParam} informs us how the thermal parameters of such a model are related to thermal parameters of each individual model.  Equally important,~\eqref{eq:def-q-agg} shows that the aggregate disturbance $\qbarAgg$ is not the average internal heat load $\bar{q}_{int}$, but incorporates an additive term ($\tilde{w}_z$) that is related to all the other states and inputs in a complex manner. This additive term is amplified when the inputs and the states of each zone vary in an asynchronous manner. 
	\end{comment}
	\begin{comment}\label{com:commentTwo}
	Although we have neglected thermal interactions in the derivation of the aggregate model, those interactions only affects additive terms. The RC network structure and the thermal parameters of the aggregate model \eqref{eq:aggParam} remain the same.
	
	Thermal interactions between the zones are modeled as an additional term to $T_z^j$ in~\eqref{eq:average-Tz-multizone}-\eqref{eq:average-Tw-multizone}, $\sum_{i\neq j}\frac{T_z^i - T_z^j}{R_z^{ij}C_z^j}$. If each $C_z^j = C_0$ for all $j$, the thermal interactions will cancel completely in the aggregate model, since $R_z^{ij} = R_z^{ji}$ by definition. If each $C_z^j$ is distinct then the aggregation error $\tilde{w}_z$ changes to $\tilde{w}^{\prime}_z$, where
	\begin{align}
	\tilde{w}^{\prime}_z & \eqdef \tilde{w}_z + \frac{1}{N_z}\sum_{i=1}^{N_z}\sum_{j=1}^{N_z} \frac{\tilde{T}_z^i-\tilde{T}_z^j}{R_z^{ij}C_z^j}.	
	\end{align}
	\end{comment}
	\section{System Identification} \label{sec:Three}
	\subsection{Problem Statement}
	We now turn to the problem of identifying the aggregate model \eqref{eq:timeInvAggModel_Tz}-\eqref{eq:timeInvAggModel_Tw} from input-output data collected from multiple zones. 
	That is, given the individual input-output data of each zone in a multi-zone building, we wish to identify the thermal parameters and internal heat load of an aggregate model which is of the same type as the individual model. Note that (i) the aggregate model takes average inputs and output which are computed from individual inputs and output, (ii) the presence of the unknown disturbances in the model, $\qbarAgg(t)$ and $\tilde{w}_w(t)$, presents a serious hurdle, especially $\qbarAgg$ since it is large; sometimes comparable to the cooling provided by the HVAC system~\cite{Kim_Braun:2016}.
	
	Our proposed approach is inspired by that in~\cite{cofbar:2018}, in which a simple dynamic model of average internal heat load is proposed by assuming it is slowly time varying. Then estimation of this unknown signal is cast as a state estimation problem. Recall that $\qbarAgg = \bar{q}_{int} + \tilde{w}_z\bar{C}_z$. In this work we assume that  $\bar{q}_{int} \gg \tilde{w}_z \bar{C}_z$, that is the average internal heat load is much larger than the scaled aggregation errors. Due to this, we adapt the following dynamic model used for $\bar{q}_{int}$ in~\cite{cofbar:2018} to $\qbarAgg$:
	\begin{align} \label{eq:dynAggDist}
		\frac{d}{dt}\qbarAgg(t) = 0
		\end{align}
	The justification for this model comes from the usual pattern of occupancy in commercial buildings, in which occupants typically enter and exit the building in bulk. This bulk transfer of people would correspond to a piece-wise constant occupant-induced heat load signal, which would consequently make its time-derivative be 0 for the most of time. Then the dynamic model for the aggregate internal heat load can be coupled with the aggregate RC network model~\eqref{eq:timeInvAggModel_Tz}-\eqref{eq:timeInvAggModel_Tw} to form the aggregate model:
	\begin{gather} 
	\begin{bmatrix} 
	\dot{\bar{T}}_z \\ \dot{\bar{T}}_w \\ \dotqbarAgg \end{bmatrix}
	=
	\begin{bmatrix}
	\frac{-1}{\bar{\tau}_{za}}+\frac{-1}{\bar{\tau}_{zw}} & \frac{1}{\bar{\tau}_{zw}}  & 1 \\
	\frac{1}{\bar{\tau}_{wz}} & \frac{-1}{\bar{\tau}_{wa}}+\frac{-1}{\bar{\tau}_{wz}} & 0 \\
	0 & 0 & 0
	\end{bmatrix}
	\begin{bmatrix}
	\bar{T}_z\\
	\bar{T}_w \\
	\qbarAgg
	\end{bmatrix}
	\nonumber\\ +
	\begin{bmatrix}
	\frac{1}{\bar{\tau}_{za}} & \bar{A}_z & \frac{1}{\bar{C}_z} \\
	\frac{1}{\bar{\tau}_{wa}} & \bar{A}_w & 0 \\
	0 & 0& 0 
	\end{bmatrix}
	\begin{bmatrix}
	\bar{T}_a\\
	\bar{\eta}_{solar}\\
	\bar{q}_{ac}\label{eq:contDynFull} 
	\end{bmatrix},	 
	\end{gather}
	where we have dropped $\tilde{w}_w$ for wall temperature is not of our interest.
	This can then be expressed in compact continuous time state space notation as,
	\begin{align} \label{eq:contDyn}
	\dot{x}(t) &= A(\theta)x(t) + B(\theta)u(t), \\
	z(t) &= Fx(t),
	\end{align}
	where the definitions of $x(t), u(t)$, $A(\theta)$, and $B(\theta)$ follow from comparison to~\eqref{eq:contDynFull}; $\theta=  [\bar{\tau}_{za},\bar{\tau}_{zw},\bar{\tau}_{wa}, \bar{\tau}_{wz},\bar{C}_z,\bar{A}_{z}\,\bar{A}_{w}]^T$ is the unknown parameter vector and $F = [0, \ 0, \ 1]$.

	\emph{The aggregate level identification problem can be posed as follows: given $N_t$ time-samples of the average measured inputs $u[k] = [\bar{T}_a[k]$, $\bar{\eta}_{solar}[k], \bar{q}_{ac}[k]]^T \in \R^3$ and the measured average output $z[k] = \bar{T}_z[k]$ for $k=0,\dots,N_t-1$, identify the unknown thermal parameter $\theta \in \R^7$, and the aggregate internal heat load signal samples $\qbarAgg[k] \in \R$ for $k=0,\dots,N_t-2$.}

    We emphasize that the average signals (defined in \eqref{eq:barVariables1}-\eqref{eq:barVariables2}) that are used as input data for the system identification problem can be computed from measurable zone-level signals. Note that $\bar{q}_{int}$ is not a input for this identification problem.

	\subsection{Proposed Identification Method}
	Our proposed method involves solving a constrained non-linear optimization problem in order to obtain estimates for the aggregate thermal parameters and the aggregate internal heat load. Since $\qbarAgg$ has been recast as a state variable, estimation of the state produces an estimate for this quantity.
	
	If the thermal parameters were already known, a Kalman filter would be adequate to estimate the state since the model is linear. However, since the thermal parameters are unknown the problem turns into a nonlinear state estimation problem. There are many available choices for non-linear state estimation, such as particle filtering, the extended Kalman filter, or moving horizon/batch estimation approaches~\cite{James:1996,anderson:1979}. In this work we choose the batch estimation approach because it allows for an easy incorporation of inequality constraints on the state estimates. Particularly, the aggregate internal heat load is enforced to be positive since it usually adds heating.
	
	To setup the batch optimization problem, the continuous time aggregate model~\eqref{eq:contDynFull} is discretized using a first order forward Euler method with a sampling time $t_s$. Additionally, process and measurement noise are included to account for modeling error. The corresponding discrete time model is then,
	\begin{align}
	x[k+1] &= x[k] + t_s\left(A(\theta)x[k] + B(\theta)u[k]\right) + G\xi[k], \\
	z[k] &= Fx[k] + \nu[k],
	\end{align}  
	where $\{\xi[k]\}_{k=0}^{N_t-2}$ and $\{\nu[k]\}_{k=0}^{N_t-1}$ are white noise sequences that capture the modeling error and sensor noise. \emph{We only let $\qbarAgg$ have process noise (i.e., $G = [0,0,1]^T$).} The reason is that since we are also identifying system parameters, adding noise to the other states would greatly increase the number of degrees of freedom and likely produce spurious results. 
	
	The batch estimation problem is now posed as:
	\begin{align} \label{alg:iden}
	&\min_{\theta,x_0,\{\xi[k]\}_{k=0}^{N_t-2}}\bigg((x_0-x^*_0)^T P_{x_0}^{-1}(x_0-x^*_0) \nonumber\\
	&\ \ \ \ \ \ \ +(\theta-\theta^*)^T P_{\Theta}^{-1} (\theta-\theta^*)	+ \lambda \sum_{k=0}^{N_t-2} |\xi[k]|\nonumber\\
	&\ \ \ \ \ \ \ +\frac{1}{r}\sum_{k=1}^{N_t-1} \nu[k]^2 + \ \alpha\sum_{k=0}^{N_t-2} \qbarAgg[k]^2\bigg), \\
	\text{s.t.} \quad &\forall \ k \in \{0,...,N_t-1\},\nonumber\\
	&\nu[k] = \bar{T}_{z}[k]-Fx[k],\nonumber\\
	&x[k+1] = x[k] + t_s\bigg(A(\theta)x[k] + B(\theta)u[k]\bigg) + G\xi[k], \nonumber\\ 
	&x[k] \in X, \quad \theta \in \Theta, \nonumber
	\end{align}
	where $x^*_0$ and $\theta^*$ are initial guesses of $x_{0}$ and $\theta$. The matrix $P_{x_0}^{-1}$ is the weighting matrix for the difference from the initial guess of initial state and $P_{\Theta}^{-1}$ is for difference from initial guess of parameters. The scalar values $\lambda, 1/r, \alpha$ represent the cost for the process, measurement noise and aggregate internal heat load, respectively. Set $X$ and $\Theta$ correspond to inequality constraints on the state and parameters based on the apriori knowledge of them.
	
	There is a penalty on aggregate heat load in the objective function because we do not want aggregate internal heat load to make up for all of the model error, giving too much degree of freedom for parameters estimation. Note we incorporate an absolute value penalty on the process noise for the aggregate internal heat load estimate. This is because we expect the derivative of $\qbarAgg$ to be sparse; similar technique is used in \cite{ZengIdentificationCPHS:2018}.

	Mathematically, this problem represents the minimization of a convex function over a non-convex set, making the overall problem a non-convex optimization problem. We solve this optimization problem with CasADi~\cite{casadi} and the NLP solver IPOPT~\cite{IPOPT_Wachter}. 
	
    \subsection{Heuristic to predict accuracy of estimated aggregate disturbance}\label{sec:heuristic}
    As is mentioned in Comment~\ref{com:commentOne}, since the proposed method identifies $\qbarAgg$ instead of $\bar{q}_{int}$, even if it performs perfectly, the estimated disturbance is likely to be different from average disturbance. The converse is also true. 
	Recalling \eqref{eq:def-q-agg}, $\qbarAgg \triangleq \bar{q}_{int} + \tilde{w}_z\bar{C}_z$, $\qbarAgg$ will differ from $\bar{q}_{int}$ more if $|\tilde{w}_z|$ is larger. To compute $|\tilde{w}_z|$, the knowledge of the thermal parameters of each zone is required (see~\eqref{eq:wTildeZ}), which is usually not available in reality. Although the exact $|\tilde{w}_z|$ is unknown, we can provide a surrogate metric for $|\tilde{w}_z|$. 
	
	Note that the sample mean and sample variance of the ``tilde" terms are defined as:
	\begin{align} 
	\mu_{p}[k] & \eqdef \frac{1}{N_z}\sum _{j=1}^{N_z}\tilde{p}^j[k],\label{eq:SampleMean}\\
	\sigma^{2}_{p}[k] & \eqdef  \frac{1}{N_z-1} \sum _{j=1}^{N_z}(\tilde{p}^j[k] - \mu_{p}[k])^2,\label{eq:SampleVariance}
	\end{align}     
	where $k$ is the discrete time step, $N_z$ is the number of zones, and $p$ represents the type of inputs or states, i.e., $p\in \{T_{a},T_z,T_w,\eta_{solar},q_{ac},q_{int}\}$. Using \eqref{eq:define_tilde}, \eqref{eq:SampleMean} is reduced to:
	\begin{align} \label{eq:ZeroMean}
	\mu_{p}[k] &= \frac{1}{N_z} \sum _{j=1}^{N_z}\tilde{p}^j[k] =\frac{1}{N_z} \sum _{j=1}^{N_z}(p^j[k]-\bar{p}[k]) \nonumber\\
	& = \frac{1}{N_z}\sum _{j=1}^{N_z}p^j[k]- \bar{p}[k] = \bar{p}[k]-\bar{p}[k] = 0.
	\end{align}
	Substituting \eqref{eq:ZeroMean} into \eqref{eq:SampleVariance}, we have:
	\begin{align}
	\sigma^{2}_{p}[k] & =  \frac{1}{N_z-1} \sum _{j=1}^{N_z}(\tilde{p}^j[k])^2.\label{eq:SampleVariance1}
	\end{align} 
	According to \eqref{eq:SampleVariance1}, large $\sigma^{2}_{p}$ implies large $\sum _{j=1}^{N_z}(\tilde{p}^j)^2$ for $p\in \{T_{a},T_z,T_w,\eta_{solar},q_{ac},q_{int}\}$. We see from \eqref{eq:wTildeZ} that $|\tilde{w}_z|$ will be large if magnitudes of these ``tilde" terms, $|\tilde{T}_z^j|$, $|\tilde{q}_{ac}^j|$, etc., are large for $j\in \{1,2,...,N_z\}$; equivalently speaking, $|\tilde{w}_z|$ will be large if $\sum _{j=1}^{N_z}(\tilde{T}_z^j)^2$, $\sum _{j=1}^{N_z}(\tilde{q}_{ac}^j)^2$, etc., are large. Therefore, $\sigma^{2}_{p}$ and $|w_z|$ are positively correlated for each $p\in \{T_{a},T_z,T_w,\eta_{solar},q_{ac},q_{int}\}$. 
	
	In summary, the variance of ``tilde" terms, $\sigma^{2}_{p}$, is a surrogate metric for aggregation error, $|\tilde{w}_z|$. Consequently, $\qbarAgg$ will differ from $\bar{q}_{int}$ more as $\sigma^{2}_{p}$ becomes larger. Since the proposed method identifies $\qbarAgg$ instead of $\bar{q}_{int}$, the estimated disturbance, $\hatqbarAgg$ is expected to differ from $\bar{q}_{int}$ more as $\sigma^{2}_{p}$ becomes larger.
	\ifx
	The sample mean and sample variance are defined as follows:
	\begin{align} 
	\mu_{x}[k] & \eqdef \frac{1}{N_z}\sum _{j=1}^{N_z}\tilde{x}^j[k],\\
	\sigma^{2}_{x}[k] & \eqdef  \frac{1}{N_z-1} \sum _{j=1}^{N_z}(\tilde{x}^j[k] - \mu_{x}[k])^2,
	\end{align}          
 	where $k$ is the discrete time step, $N_z$ is the number of zones, and $x$ can be any ``tilde'' term in~\eqref{eq:wTildeZ}. There are 6 variances of ``tilde" terms in total: $\sigma^{2}_{T_a},\ \sigma^{2}_{T_z}$, $\sigma^{2}_{T_w}$, $\sigma^{2}_{q_{ac}}$, $\sigma^{2}_{\eta_{solar}}$, and $\sigma^{2}_{q_{int}}$. Now we explain why the magnitude of these sample variances indicates the magnitude of $|\tilde{w}_z|$.
	
	We see from \eqref{eq:wTildeZ} that $|\tilde{w}_z|$ will be larger when the ``tilde" terms, $|\tilde{T}_z^j|$, $|\tilde{q}_{ac}^j|$, etc., are larger. Since the ``tilde" terms are defined as the deviation from their zonal average (see \eqref{eq:define_tilde}), their magnitude can be . This deviation from average can be reflected by its variance.  
	By backtracking this analysis, we know that if the variances of the ``tilde" terms are large at time $t$, $|\tilde{w}_z(t)|$ is expected to be large, and thus $\qbarAgg(t)$ will differ significantly from $\bar{q}_{int}(t)$. 
    Since the variances of the ``tilde" terms can be estimated from the their sample variances ($\sigma^{2}_{T_a},\ \sigma^{2}_{T_z}$, $\sigma^{2}_{T_w}$, $\sigma^{2}_{q_{ac}}$, $\sigma^{2}_{\eta_{solar}}$, $\sigma^{2}_{q_{int}}$), larger the sample variances are, more different the $\qbarAgg$ and $\bar{q}_{int}$ will be.
	\fi
	\section{Evaluation with simulation data and real building data} \label{sec:Four}
	We present results for two datasets (i) data generated from an open loop simulation of a multi-zone building model (``virtual building'') and (ii) data collected from a testbed located in the Oak Ridge National Laboratory while the building operates under normal closed loop operation. The open loop scenario is included to illustrate some of the analytical concepts derived.
	
	\subsection{Evaluation with simulation data}\label{sec:eval-sim}
	The virtual building has $5$ independent zones, each modeled by a 2R2C network model~\eqref{eq:Ind_model_Tz}-\eqref{eq:Ind_model_Tw} and is aggregated according to~\eqref{eq:average-Tz-multizone}-\eqref{eq:average-Tw-multizone}. The simulation data is generated from a 2 day open loop simulation and is aggregated according to the aggregation method~\eqref{eq:barVariables1}-\eqref{eq:barVariables2}. This data is used by the proposed identification method to estimate the parameters of an aggregate  2R2C network model~\eqref{eq:timeInvAggModel_Tz}-\eqref{eq:timeInvAggModel_Tw} along with the aggregate internal heat load.
	\begin{table}
		\centering
		\caption{Parameter estimates from open loop simulation (virtual building).}
		\setlength{\arrayrulewidth}{0.03cm}
		\label{tab:simParam}
		\begin{tabular}{c c c c}
			\hline
			Parameter & Estimate & True Value & units \\
			\hline
			$\bar{\tau}_{za}$ & 0.7652 & 0.7899 &hour(s) \\
			$\bar{\tau}_{zw}$ & 0.5919 & 0.5869 &hour(s) \\
			$\bar{C}_{z}$ & 0.7050 & 0.7147 &$kWh/^\circ C$ \\
			$\bar{A}_z$ & 0.7884 & 0.5700 &$^\circ C m^2/kWh$ \\ 
			$\bar{\tau}_{wa}$ & 24.5098 & 19.3798 &hour(s) \\
			$\bar{\tau}_{wz}$ & 2.6795 & 2.8441 &hour(s) \\
			$\bar{A}_w$ & 4.2955 & 4.5537 &$^\circ C m^2/kWh$ \\
			\hline
		\end{tabular}
	\end{table}

\begin{figure}
		\centering
		\includegraphics[width =1\columnwidth,height = 0.7\columnwidth]{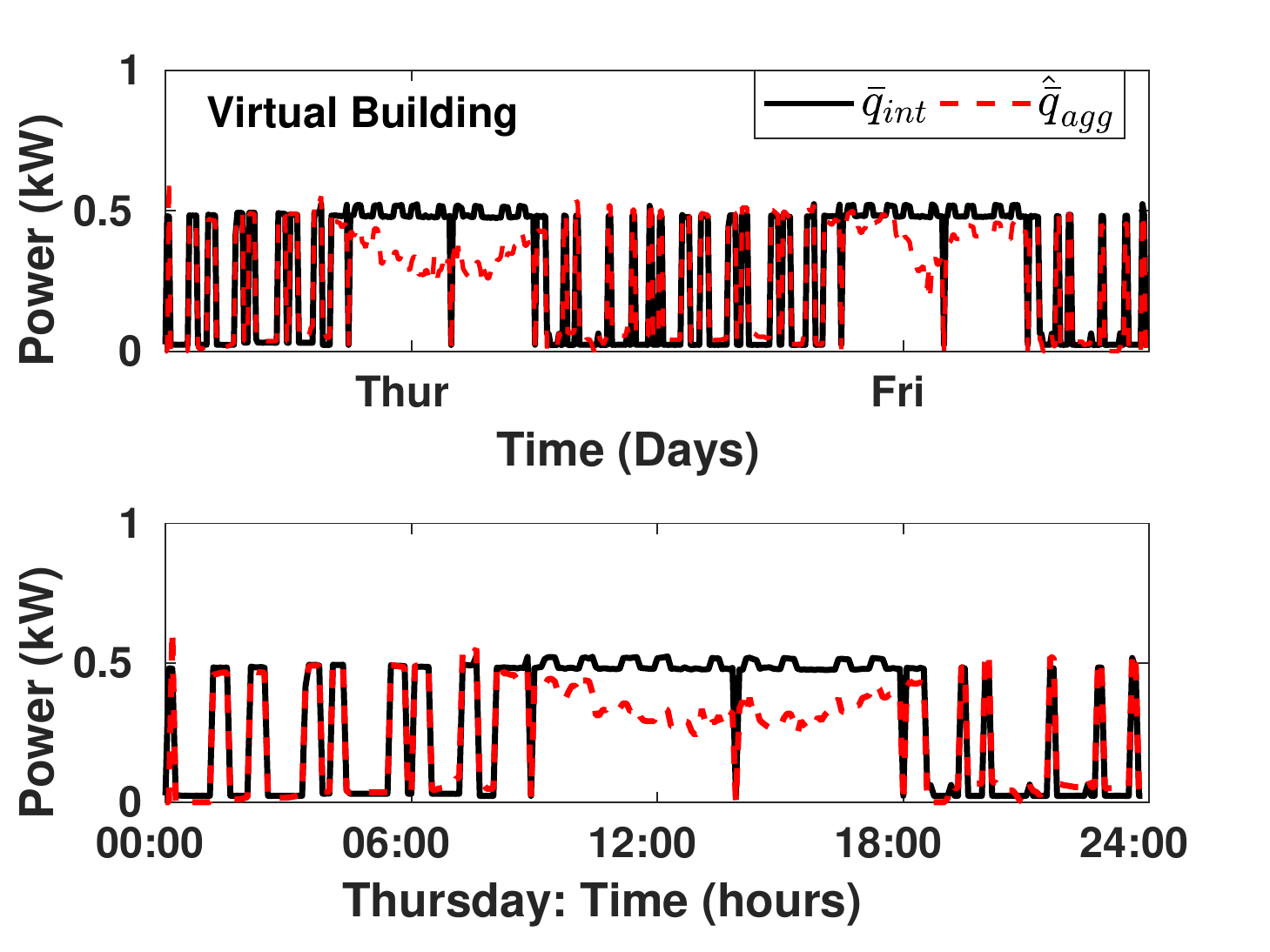}
		\caption{Aggregate disturbance estimate $\hatqbarAgg$ and average internal disturbance $\bar{q}_{int}$ for the simulated virtual building.}
		\label{fig:simResults}
	\end{figure}

	 Table~\ref{tab:simParam} shows the parameter estimation results along with their true values. As one can see from the table, the parameters are estimated quite accurately. The estimated aggregate internal heat load $\hatqbarAgg$ and the true average internal heat load $\bar{q}_{int}$ are shown in Figure~\ref{fig:simResults}. We can see obvious differences between $\hatqbarAgg$ and $\bar{q}_{int}$.
	
	\begin{figure}
		\centering
		\includegraphics[width =1\columnwidth,height = 0.425\columnwidth]{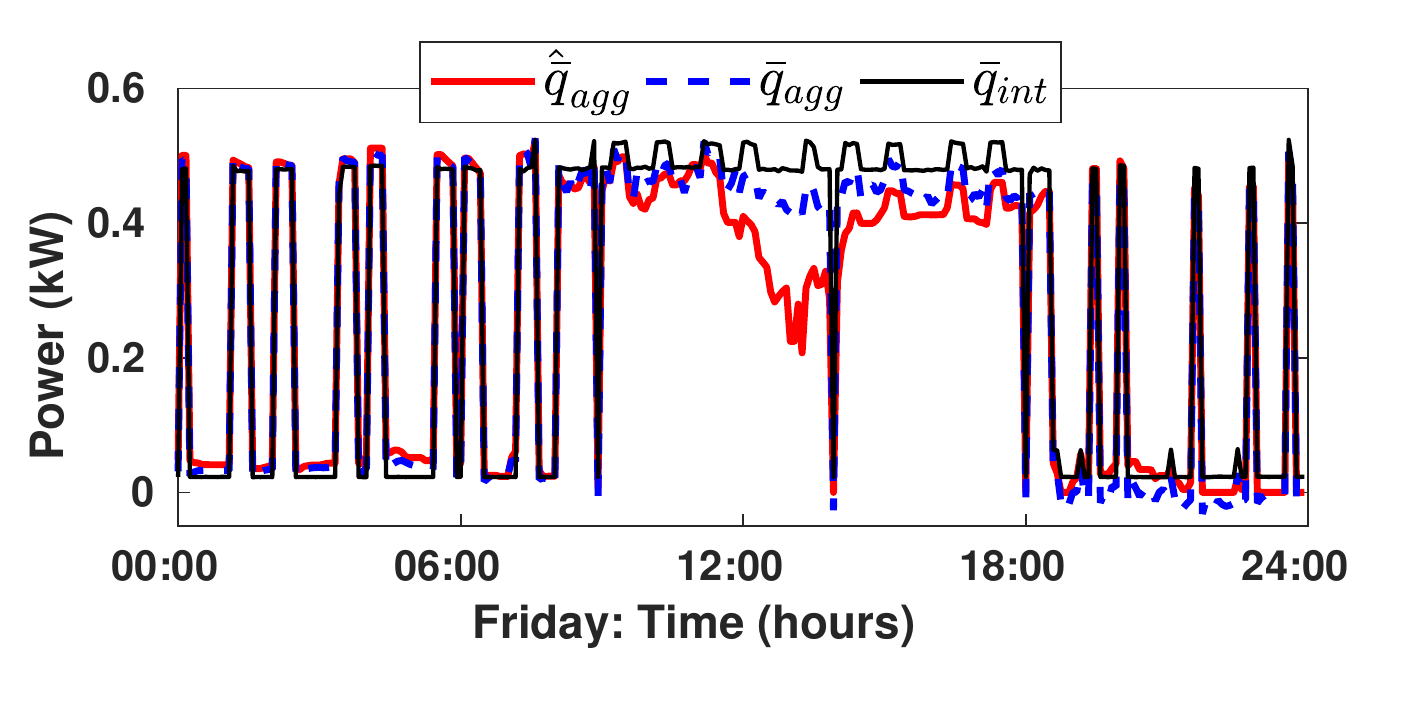}
		\caption{Comparison of $\qbarAgg$, its estimate $\hatqbarAgg$, and $\bar{q}_{int}$.}
		\label{fig:wTildeSim}
	\end{figure}
	Since all parameters and measurements in the simulation are known, we can compute $w_z$ to obtain the true aggregate internal heat load, $\qbarAgg$. Figure~\ref{fig:wTildeSim} shows the estimated aggregate internal heat load $\hatqbarAgg$, the true aggregate internal heat load $\qbarAgg$, and the true average internal heat load  $\bar{q}_{int}$. Recall from~\eqref{eq:def-q-agg} that the aggregate internal heat loads differs from the average heat load by $\tilde{w}_z(t){C}_z$. In fact this difference is clearly seen in Figure~\ref{fig:wTildeSim}. Besides, we can see the identification method is trying to estimate the aggregate internal heat load ($\qbarAgg$), not average internal heat load ($\bar{q}_{int}$). \emph{Therefore, even if the method performs perfectly there will be some non-vanishing difference between the estimated quantity, $\hatqbarAgg$, and the true average internal heat load, $\bar{q}_{int}$.} This observation should be kept in mind in interpreting the estimation results. Otherwise the estimation may appear to be poorer than it actually is.
	\subsection{Evaluation with ORNL building data}
\begin{figure}
		\centering
		\includegraphics[width=0.6\columnwidth]{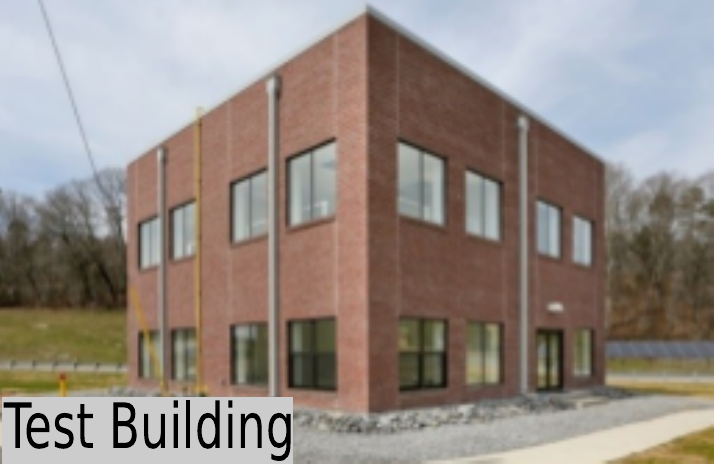}
		\caption{ORNL test building}
		\label{fig:ornlBuild}
	\end{figure}

	The data for this evaluation is collected from a test building located at Oak Ridge National Laboratory (ORNL), Oak Ridge, TN, shown in Figure~\ref{fig:ornlBuild}.
	\subsubsection{Data Collection}
	 Data from 9-21-2018 to 10-03-2018 is used for identification. The building is unoccupied during this time. To mimic the effects of occupancy ($\bar{q}_{int}$), space heaters placed in each room are turned on and off synchronously. Since the rating of each space heater is known, we have the knowledge of ${q}_{int}^j$ for all $jth$ zone and therefore the knowledge of $\bar{q}_{int}$. Additionally, the VAV for each zone is controlled to maintain the temperature of each zone within a preset range during daily operation. During nights, the VAVs are shut off.

\begin{figure}
		\centering
		\includegraphics[width=1\linewidth]{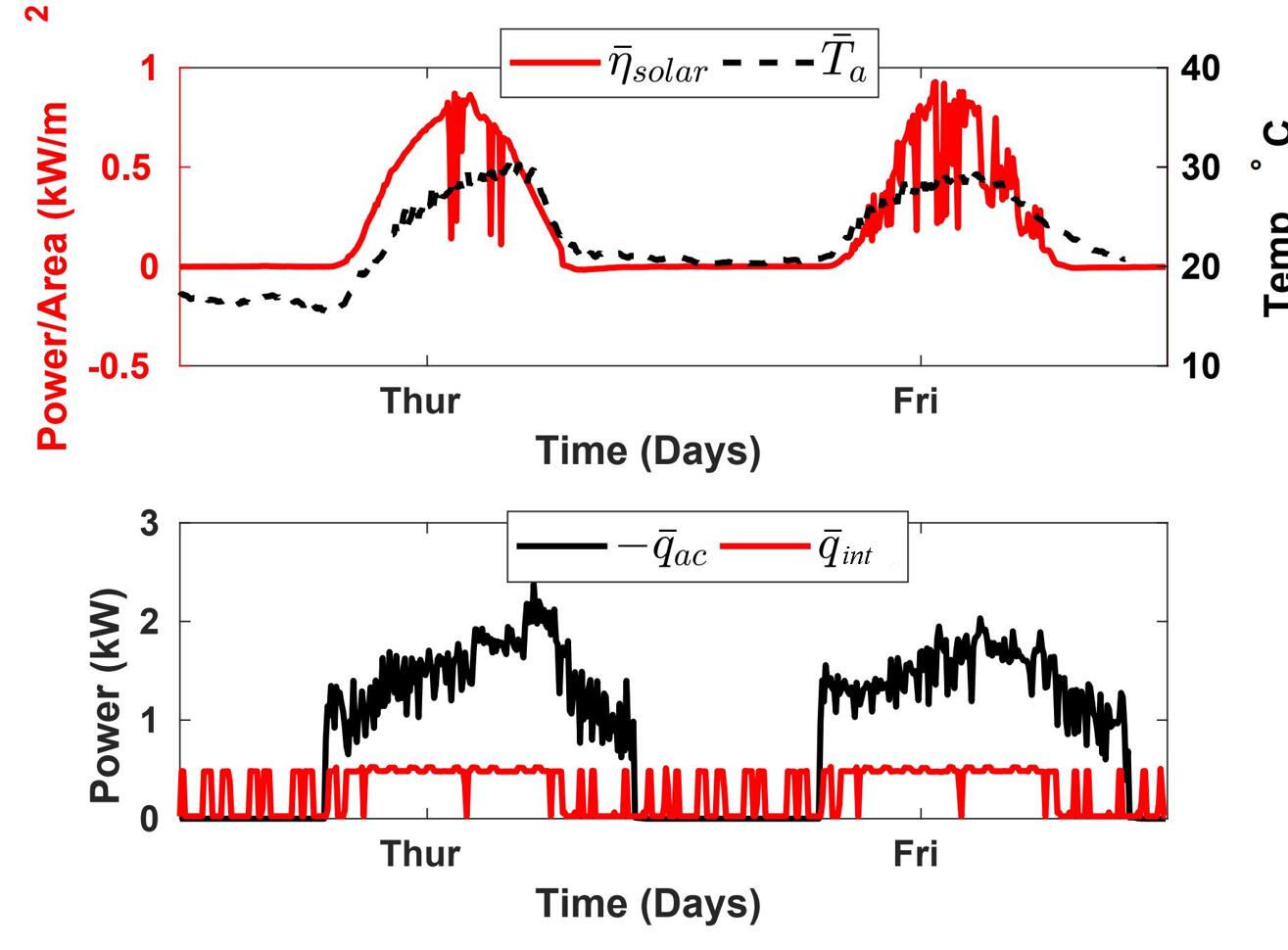}
		\caption{Aggregated input sequences for the ORNL test building.}
		\label{fig:inpORNL}
	\end{figure}

	The inputs to the system ($\bar{q}_{ac}$, $\bar{\eta}_{solar}$,$\bar{T}_a$ and $\bar{q}_{int}$) are shown in Figure~\ref{fig:inpORNL}. These inputs are computed from measurements from individual zones using equations~\eqref{eq:barVariables1}-\eqref{eq:barVariables2}. Since we have assumed that $\qbarAgg$ is primarily composed of $\bar{q}_{int}$, we compare the estimation results of $\qbarAgg$ to the measured $\bar{q}_{int}$.
	
	\subsubsection{In-sample Results}
	We train our model on $8$ days of collected training data. Table~\ref{tab:ornlParam} presents the estimated aggregate thermal parameters. Unlike the virtual building case study, there is no ground truth to compare them to.  Therefore we resort to cross validation, which is discussed in Section~\ref{sec:eval-ornl-oos}.
		
	\begin{table}
		\centering
		\caption{Parameter estimates from closed-loop data collected from ORNL test building.}
		\setlength{\arrayrulewidth}{0.03cm}
		\label{tab:ornlParam}
		\begin{tabular}{c c c}
			\hline
			Parameter & Estimate & units \\
			\hline
			$\bar{\tau}_{za}$ & 2.08 & hour(s) \\
			$\bar{\tau}_{zw}$ & 0.2761 & hour(s) \\
			$\bar{C}_{z}$ & 0.1019 & $kWh/^\circ C$ \\
			$\bar{A}_z$ & 10.2 & $^\circ C m^2/kWh$ \\ 
			$\bar{\tau}_{wa}$ & 200 & hour(s) \\
			$\bar{\tau}_{wz}$ & 3.4722 & hour(s) \\
			$\bar{A}_w$ & 0.7164 & $^\circ C m^2/kWh$ \\
			\hline
		\end{tabular}
	\end{table}
	
	Figure~\ref{fig:ornlDistResults} shows the estimated aggregate internal heat load $\hatqbarAgg$ and average internal heat load $\bar{q}_{int}$. To avoid clutter we only present $2$ days of data. The estimated aggregate internal heat load is consistent with the average internal heat load during nighttime while is sometimes significantly different from the average internal heat load in daytime. Unlike the simulation case, the true aggregate internal heat load $\qbarAgg$ cannot be computed due to lack of knowledge of parameters of each zone. Thus it is hard to determine how close is the $\hatqbarAgg$ to $\qbarAgg$, i.e., how well the estimation of aggregate internal heat load is.

        \subsubsection{Use of heuristic to predict internal load estimation accuracy}\label{Sec:heuristic-application}
\begin{figure}
		\centering
		\includegraphics[width=1\linewidth]{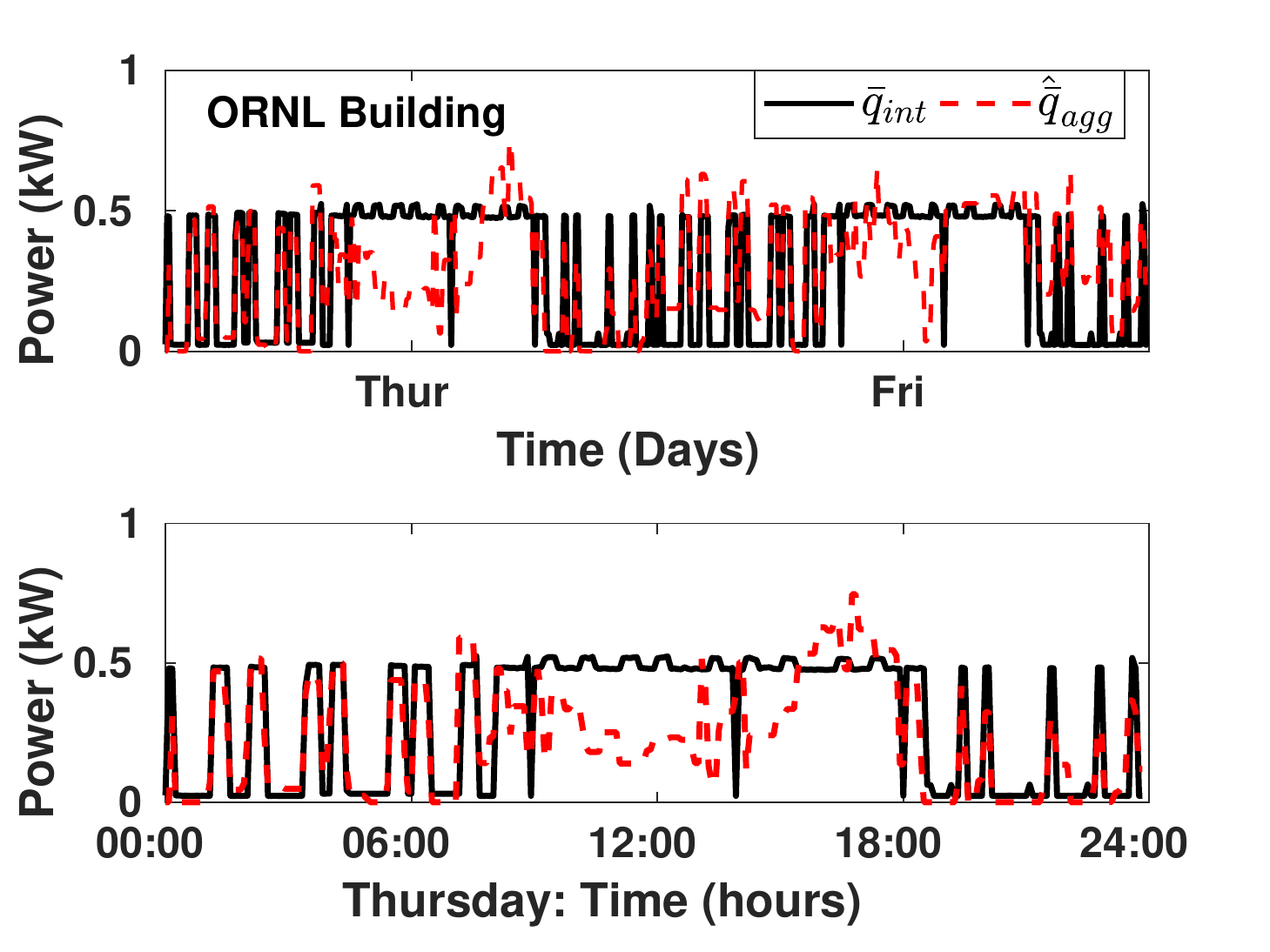}
		\caption{Evaluation with building data: estimated aggregate internal disturbance $\hatqbarAgg$ and measured average internal disturbance $\bar{q}_{int}$. Top: results for 2 days; bottom: zoomed in for one day.}
		\label{fig:ornlDistResults}
		\includegraphics[width=1\columnwidth]{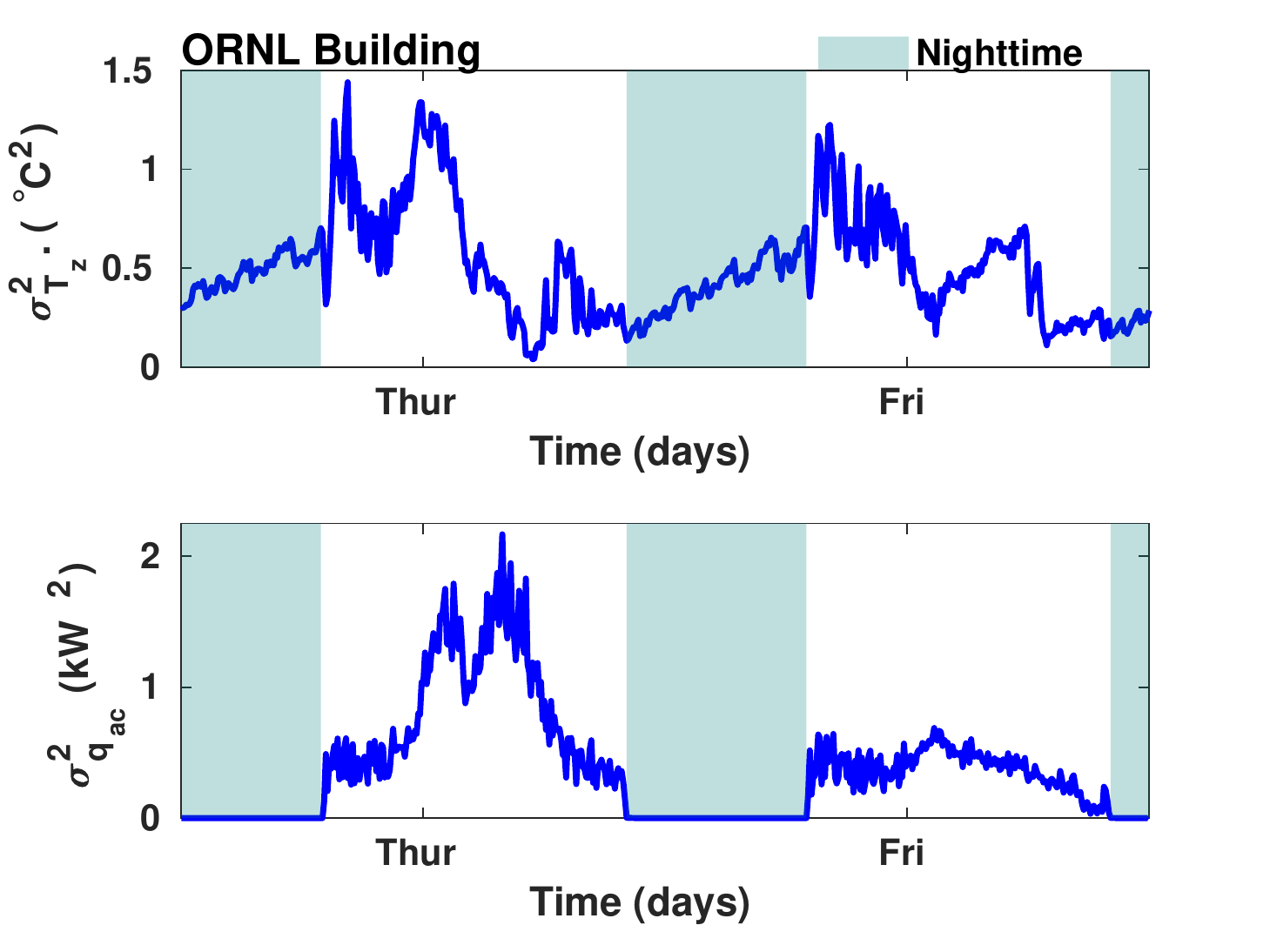}
		\caption{Building data: sample variance in Thursday and Friday to illustrate the level of asynchronicity of terms that contribute to $\tilde{w}_z$. Top: sample variance of $\tilde{T}_{z}^j$ over zones; bottom: sample variance of $\tilde{q}_{ac}^j$ over zones, where $j$ is zone index.} 
		\label{fig:varInpPlot_Insample}
	\end{figure}

        We now apply the heuristic described in Section~\ref{sec:heuristic}. We only compute sample variances of $\tilde{T}_z^j$ and $\tilde{q}_{ac,k}^j$ because the other terms incorporated in $\tilde{w}_z$ are either synchronous across zones (such as the ambient temperature), unmeasurable (such as the wall temperature), or their measurement resolutions may not allow for their computation (such as the solar irradiance). The sample variances are shown in Figure~\ref{fig:varInpPlot_Insample}.
         
        As we can see, the sample variances are considerably larger during the day, which is exactly when $|\hatqbarAgg-\bar{q}_{int}|$ is also at its largest (Figure~\ref{fig:ornlDistResults}). The heuristic thus explains - at least partly -  why the differences between $\hatqbarAgg$ and $\bar{q}_{int}$ are significant during nighttime and small during daytime.

        Furthermore, the sample variances in Thursday daytime is larger than them in Friday daytime, meaning the  $|\hatqbarAgg-\bar{q}_{int}|$ in Thursday daytime should be larger than it in Friday daytime, which is exactly what happens; see Figure~\ref{fig:ornlDistResults}. Thus, one can use the heuristic to assess at which time periods the load estimates are likely to be less representative of the true load than at other times.

	\subsubsection{Out of sample (OOS) results}\label{sec:eval-ornl-oos}
\begin{figure}[h]
		\centering
		\includegraphics[width=1\columnwidth]{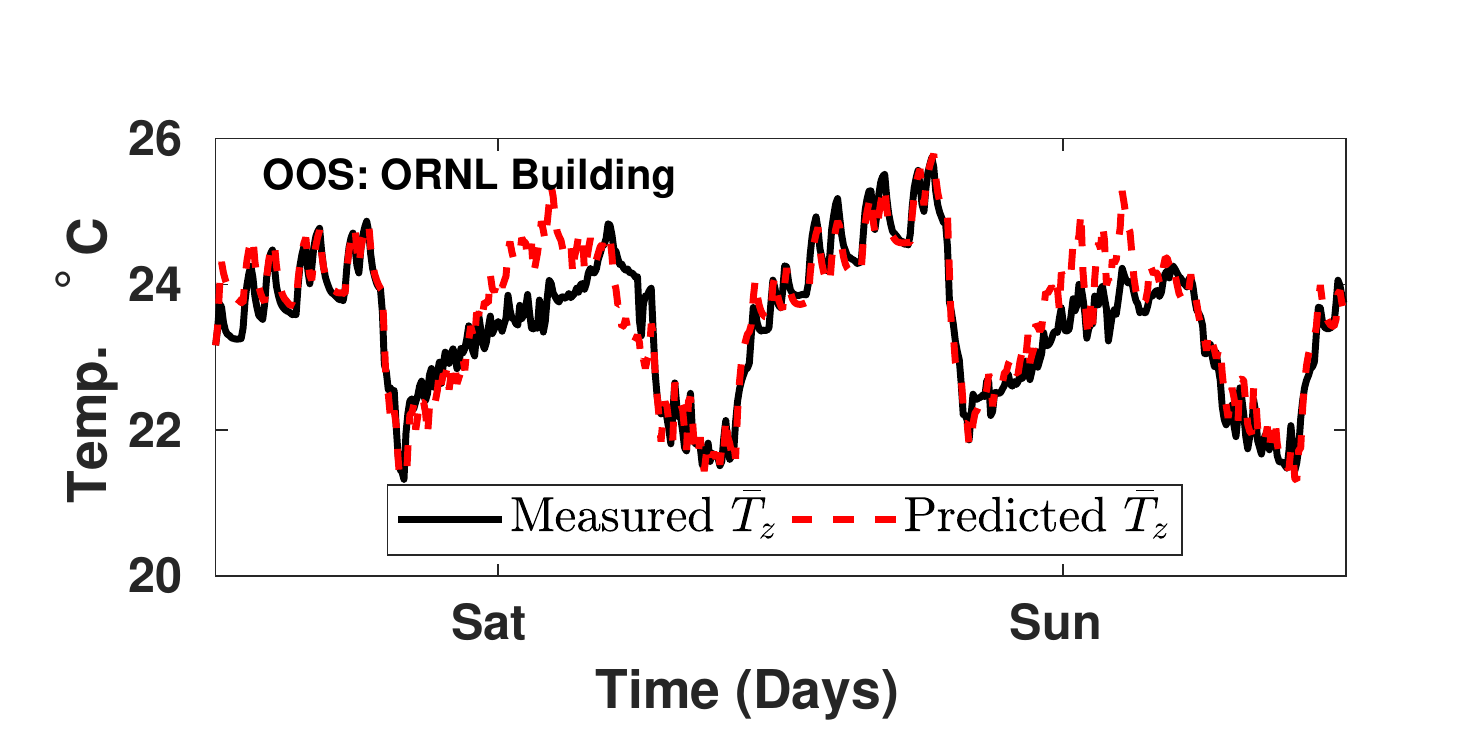}
		\caption{Out of sample aggregate zone temperature ($\bar{T}_z$) prediction results for the ORNL building using the estimated aggregate RC network model.}
		\label{fig:oosTempPred}
	    \includegraphics[width=1\columnwidth]{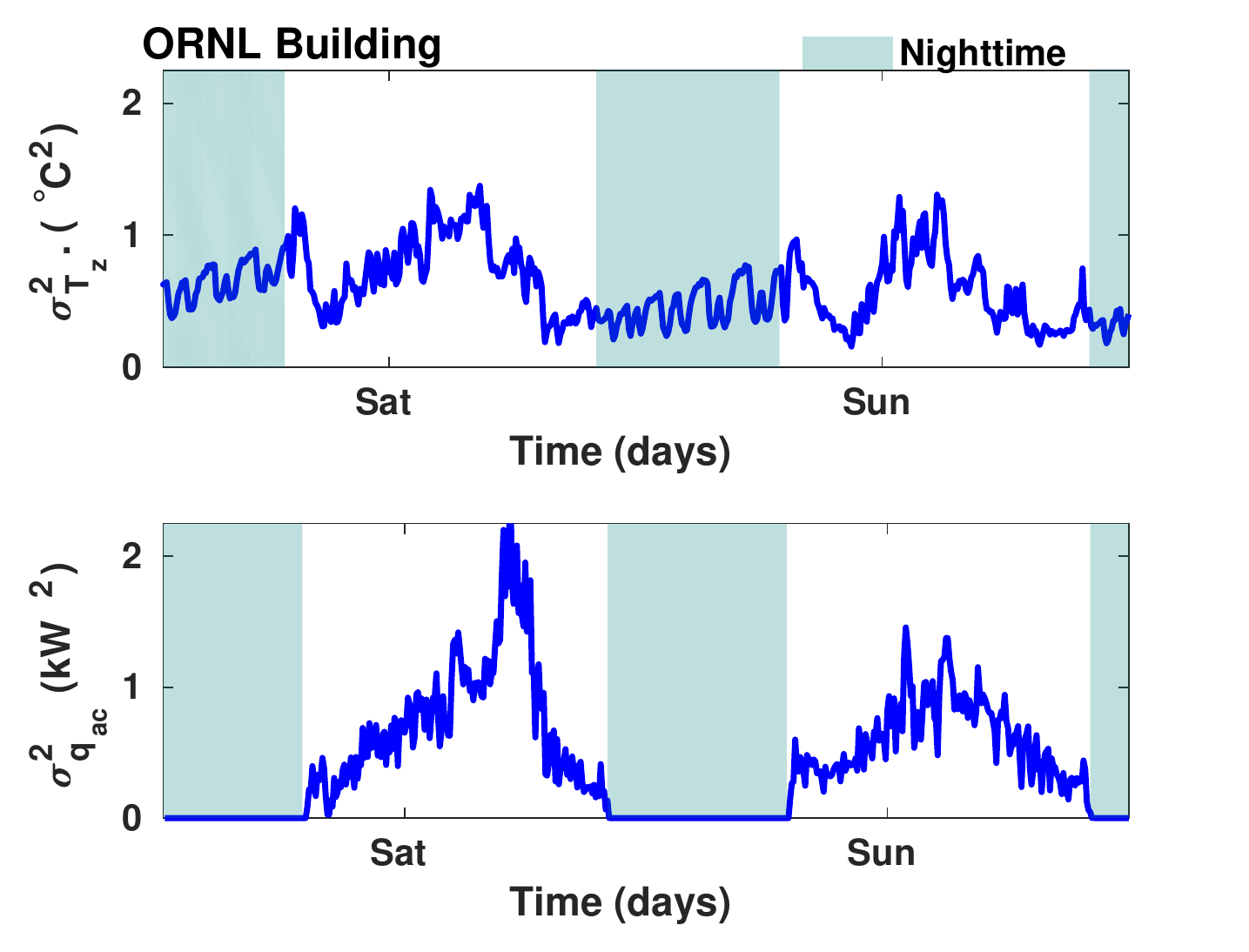}
		\caption{Building data: sample variance in Saturday and Sunday to illustrate the level of asynchronicity of terms that contribute to $\tilde{w}_z$. Top: sample variance of $\tilde{T}_{z}^j$ over zones; bottom: sample variance of $\tilde{q}_{ac}^j$ over zones, where $j$ is zone index.}
		\label{fig:varInpPlot_Outsample}
	\end{figure}

	The thermal parameters estimated (Table~\ref{tab:ornlParam}) from the training dataset are used  to predict aggregated zone temperature for a testing dataset that is distinct from the training data. Since we only have average internal heat load, we utilize it along with the estimated parameters and known inputs to compute these out-of-sample predictions. The results are shown in Figure~\ref{fig:oosTempPred}. The successful out-of-sample prediction suggests that the thermal parameters identified are reasonably accurate. 
	
	 As is discussed in Section~\ref{sec:heuristic}, since we are using average internal heat load $\bar{q}_{int}$ instead of aggregate internal heat load $\qbarAgg$, we would expect larger temperature prediction errors when $|\qbarAgg-\bar{q}_{int}|$ is larger, which is the time when the inputs and states are more asynchronous. Similar to what we did in Section~\ref{Sec:heuristic-application}, we plot sample variances of $\tilde{T}_{z}^j$ and $\tilde{q}_{ac}^j$ over zones to indicate the level of asynchronicities of them. It is observed that the average zone temperature prediction error is typically larger when $\tilde{T}_z^j$ and $\tilde{q}_{ac}^j$ are more asynchronous, which accords with our expectation.
	\section{Conclusion} \label{sec:conclusion}
	We derive an aggregate (single-zone equivalent) model of a multi-zone building. Our aggregation method shows that unless all zones in the building are identical (i.e synchronous inputs or identical thermal parameters) the aggregate model will be affected by time varying additive terms. These additive terms act as an additional heating/cooling source to the model. In fact, in~\cite{braun2002inverse} the authors apply similar definitions of the aggregate input/output signals to collected data from a multi-zone building. When the cooling load is predicted, they observe that model errors were most sensitive to the magnitude of the internal heat loads. The authors in~\cite{braun2002inverse} provide one explanation for this. However, another possible explanation is that the additive terms due to aggregation were not accounted for. 
	
	The identification technique we propose can identify the parameters of the derived aggregate model and the aggregate internal heat load. The aggregate internal heat load is fundamentally distinct from the average internal heat load, so that even with careful instrumentation a ground truth cannot be established. The heuristic we proposed is useful in this context; it allows one to assess at time time periods the internal load estimates are likely to be error prone. The identification method is then tested on data from a unique testbed in ORNL which allows the internal heat load to be measured. The results corroborate the effectiveness of our method and the utility of the heuristic. 
	
	There are several avenues for future work, such as examination of the effect of latent heat on estimation accuracy, analysis of data quality required to ensure high quality estimation~\cite{CaiOptimizing:ACC:2016}, and methods for identification of a multi-zone building model without aggregation.

    \section*{Acknowledgment}
    The research reported here has been partially supported by the NSF through awards 1463316, 1646229, 1934322, and DOE GMLC grant titled ``virtual batteries". The authors would like to thank Naren Raman and Bo Chen for help with the Casadi software.

    \biboptions{sort&compress}


\begin{thebibliography}{10}
    	\expandafter\ifx\csname url\endcsname\relax
    	\def\url#1{\texttt{#1}}\fi
    	\expandafter\ifx\csname urlprefix\endcsname\relax\def\urlprefix{URL }\fi
    	\expandafter\ifx\csname href\endcsname\relax
    	\def\href#1#2{#2} \def\path#1{#1}\fi
    	
    	\bibitem{afram2014theory}
    	A.~Afram, F.~Janabi-Sharifi, Theory and applications of {HVAC} control
    	systems--a review of model predictive control ({MPC}), Building and
    	Environment 72 (2014) 343--355.
    	
    	\bibitem{Ma:2012}
    	Y.~{Ma}, A.~{Kelman}, A.~{Daly}, F.~{Borrelli}, Predictive control for energy
    	efficient buildings with thermal storage: Modeling, stimulation, and
    	experiments, IEEE Control Systems Magazine 32~(1) (2012) 44--64.
    	
    	\bibitem{Morris:94}
    	F.~B. Morris, J.~E. Braun, S.~J. Treado, Experimental and simulated performance
    	of optimal control of building thermal storage, {ASHRAE} transactions 100~(1)
    	(1994) 402--414.
    	
    	\bibitem{Sun:2013}
    	Y.~Sun, S.~Wang, F.~Xiao, D.~Gao, Peak load shifting control using different
    	cold thermal energy storage facilities in commercial buildings: A review,
    	Energy conversion and management 71 (2013) 101--114.
    	
    	\bibitem{haomidbarmey:2012}
    	H.~Hao, T.~Middelkoop, P.~Barooah, S.~Meyn, How demand response from commercial
    	buildings will provide the regulation needs of the grid, in: 50th Annual
    	Allerton Conference on Communication, Control and Computing, 2012, invited
    	paper.
    	
    	\bibitem{Lin:2015}
    	Y.~{Lin}, P.~{Barooah}, S.~{Meyn}, T.~{Middelkoop}, Experimental evaluation of
    	frequency regulation from commercial building hvac systems, IEEE Transactions
    	on Smart Grid 6~(2) (2015) 776--783.
    	
    	\bibitem{WANG2019109405}
    	Z.~Wang, Y.~Chen, Data-driven modeling of building thermal dynamics:
    	Methodology and state of the art, Energy and Buildings 203 (2019) 109405.
    	
    	\bibitem{Fux:2014}
    	S.~F. Fux, A.~Ashouri, M.~J. Benz, L.~Guzzella, {EKF} based self-adaptive
    	thermal model for a passive house, Energy and Buildings 68, Part C (2014) 811
    	-- 817.
    	
    	\bibitem{Bacher:2011}
    	P.~Bacher, H.~Madsen, Identifying suitable models for the heat dynamics of
    	buildings, Energy and Buildings 43~(7) (2011) 1511--1522.
    	
    	\bibitem{Wang:2006}
    	S.~Wang, X.~Xu, Parameter estimation of internal thermal mass of building
    	dynamic models using genetic algorithm, Energy Conversion and Management
    	47~(13–14) (2006) 1927 -- 1941.
    	
    	\bibitem{Andersen:2000}
    	K.~K. Andersen, H.~Madsen, L.~H. Hansen, Modelling the heat dynamics of a
    	building using stochastic differential equations, Energy and Buildings 31~(1)
    	(2000) 13 -- 24.
    	
    	\bibitem{Kim_Braun:2016}
    	D.~Kim, J.~Cai, K.~B. Ariyur, J.~E. Braun, System identification for building
    	thermal systems under the presence of unmeasured disturbances in closed loop
    	operation: Lumped disturbance modeling approach, Building and Environment 107
    	(2016) 169 -- 180.
    	
    	\bibitem{cofbar:2018}
    	A.~Coffman, P.~Barooah, Simultaneous identification of dynamic model and
    	occupant-induced disturbance for commercial buildings, Building and
    	Environment 128 (2018) 153--160.
    	
    	\bibitem{ZengSimultaneousHPB:2018}
    	T.~Zeng, J.~Brooks, P.~Barooah, Simultaneous identification of building dynamic
    	model and disturbance using sparsity-promoting optimization, in: 5th
    	International Conference on High Performance Buildings, 2018, pp. 1--10.
    	
    	\bibitem{PahwaThesis:1983}
    	A.~Pahwa, Physical-stochastic modeling of power system loads: modeling and
    	system identification of a residential air conditioning system, Ph.D. thesis,
    	{Texas A\&M University} (December 1983).
    	
    	\bibitem{ZengIdentificationCPHS:2018}
    	T.~Zeng, P.~Barooah, Identification of network dynamics and disturbance for a
    	multi-zone building, in: 2nd {IFAC} Conference on Cyber-Physical and Human
    	Systems ({CPHS'18}), 2018.
    	
    	\bibitem{ThibaultPean:2018}
    	T.~{Pean}, J.~{Salom}, R.~{Costa-Castelloó}, Configurations of model
    	predictive control to exploit energy flexibility in building thermal loads,
    	in: 2018 IEEE Conference on Decision and Control (CDC), 2018, pp. 3177--3182.
    	
    	\bibitem{Ma:2014}
    	J.~Ma, S.~J. Qin, T.~Salsbury, Application of economic {MPC} to the energy and
    	demand minimization of a commercial building, Journal of Process Control
    	24~(8) (2014) 1282 -- 1291, economic nonlinear model predictive control.
    	
    	\bibitem{patel2018economic}
    	N.~R. Patel, J.~B. Rawlings, M.~J. Ellis, M.~J. Wenzel, R.~D. Turney, An
    	economic model predictive control framework for distributed embedded battery
    	applications (2018).
    	
    	\bibitem{energies:MPC}
    	G.~Serale, M.~Fiorentini, A.~Capozzoli, D.~Bernardini, A.~Bemporad, Model
    	predictive control {MPC} for enhancing building and hvac system energy
    	efficiency: Problem formulation, applications and opportunities, Energies
    	11~(3) (2018).
    	
    	\bibitem{Deng:2014_Automatica}
    	K.~Deng, S.~Goyal, P.~Barooah, P.~G. Mehta, Structure-preserving model
    	reduction of nonlinear building thermal models, Automatica 50~(4) (2014) 1188
    	-- 1195.
    	
    	\bibitem{GuoIdentification:CDC:19}
    	Z.~Guo, A.~R. Coffman, J.~Munk, P.~Im, P.~Barooah, Identification of aggregate
    	building thermal dynamic model and unmeasured internal heat load from data,
    	in: {IEEE} Conference on Decision and Control, 2019, accepted.
    	
    	\bibitem{James:1996}
    	D.~G. Robertson, J.~H. Lee, J.~B. Rawlings, A moving horizon-based approach for
    	least-squares estimation, AIChE Journal 42~(8) (1996) 2209--2224.
    	
    	\bibitem{anderson:1979}
    	B.~Anderson, J.~Moore, {Optimal Filtering}, Prentice-Hall, Englewood Cliffs,
    	NJ, 1979.
    	
    	\bibitem{casadi}
    	J.~Andersson, J.~{\AA}kesson, M.~Diehl, Casadi: A symbolic package for
    	automatic differentiation andoptimal control, in: S.~Forth, P.~Hovland,
    	E.~Phipps, J.~Utke, A.~Walther (Eds.), Recent Advances in Algorithmic
    	Differentiation, Springer Berlin Heidelberg, 2012, pp. 297--307.
    	
    	\bibitem{IPOPT_Wachter}
    	A.~W{\"a}chter, L.~B, On the implementation of an interior-point filter
    	line-search algorithm for large-scale nonlinear programming, Mathematical
    	Programming 106~(1) (2006) 25--57.
    	
    	\bibitem{braun2002inverse}
    	J.~E. Braun, N.~Chaturvedi, An inverse gray-box model for transient building
    	load prediction, HVAC\&R Research 8~(1) (2002) 73--99.
    	
    	\bibitem{CaiOptimizing:ACC:2016}
    	J.~Cai, D.~Kim, J.~E. Braun, J.~Hu, Optimizing zone temperature setpoint
    	excitation to minimize training data for data-driven dynamic building models,
    	in: American control conference, 2016.
    	
    \end{thebibliography}
    \end{document}